\documentclass[pdflatex,sn-mathphys-num]{sn-jnl}

\usepackage{graphicx}%
\usepackage{multirow}%
\usepackage{amsmath,amssymb,amsfonts}%
\usepackage{amsthm}%
\usepackage{mathrsfs}%
\usepackage[title]{appendix}%
\usepackage{xcolor}%
\usepackage{textcomp}%
\usepackage{manyfoot}%
\usepackage{booktabs}%
\usepackage{algorithm}%
\usepackage{algorithmicx}%
\usepackage{algpseudocode}%
\usepackage{listings}%
\usepackage{longtable}

\usepackage[draft,authormarkup=none]{changes}
\definechangesauthor[color=red]{JR}
\definechangesauthor[color=blue]{SD}

\theoremstyle{thmstyleone}%
\theoremstyle{thmstyletwo}%

\theoremstyle{thmstylethree}%

\begin{document}

\title[Characterizing asymmetric and bimodal long-term financial return distributions through quantum walks]{Characterizing asymmetric and bimodal long-term financial return distributions through quantum walks}


\author*[1,2]{\fnm{Stijn} \sur{De Backer}}\email{stijn.debacker@ugent.be}

\author[1,2]{\fnm{Luis} \spfx{E.C.} \sur{Rocha}}\email{luis.rocha@ugent.be}

\author[1]{\fnm{Jan} \sur{Ryckebusch}}\email{jan.ryckebusch@ugent.be}

\author[2]{\fnm{Koen} \sur{Schoors}}\email{koen.schoors@ugent.be}

\affil*[1]{\orgdiv{Department of Physics and Astronomy}, \orgname{Ghent University}, \orgaddress{\street{Proeftuinstraat 86}, \city{Ghent}, \postcode{9000}, \country{Belgium}}}

\affil[2]{\orgdiv{Department of Economics}, \orgname{Ghent University}, \orgaddress{\street{Sint-Pietersplein 5}, \city{Ghent}, \postcode{9000}, \country{Belgium}}}


\abstract{The analysis of logarithmic return distributions defined over large time scales is crucial for understanding the long-term dynamics of asset price movements. For large time scales of the order of two trading years, the anticipated Gaussian behavior of the returns often does not emerge, and their distributions often exhibit a high level of asymmetry and bimodality. These features are inadequately captured by the majority of classical models to address financial time series and return distributions. In the presented analysis, we use a model based on the discrete-time quantum walk to characterize the observed asymmetry and bimodality. The quantum walk distinguishes itself from a classical diffusion process by the occurrence of interference effects, which allows for the generation of bimodal and asymmetric probability distributions. By capturing the broader trends and patterns that emerge over extended periods, this analysis complements traditional short-term models and offers opportunities to more accurately describe the probabilistic structure underlying long-term financial decisions.}




\maketitle

\section{Introduction}
\label{Section:Introduction}
The analysis of financial time series is fundamental to understanding market behavior, risk assessment, and economic trends, making it a key element of research in both finance and econophysics. The first model to describe the dynamics of financial time series was developed by Bachelier in 1900~\citep{Bachelier1900}, in which the movements of asset prices were governed by a classical random walk as underlying stochastic process. This model was unable to capture fat-tailed return distributions and the corresponding extreme price changes. Bachelier noted these extreme price fluctuations but categorized them as outliers that did not deserve further attention~\citep{Mandelbrot1997variation}. In 1963, Mandelbrot proposed Lévy stable distributions as a model for financial return distributions, which included the possibility to address fat tails~\citep{Mandelbrot1963variationcertain}. Since then, a plethora of models for financial time series and return distributions has been proposed~\citep{ziemann2021physics, BlackScholes1973pricing, Gabaix2009power, Gopikrishnan1998inverse, gopikrishnan1999scaling, yuan2018cev, rogers1997arbitrage, rostek2013note, kou2002jump, merton1976option, hanson2002jump, cont1997scaling, eberlein1995hyperbolic, barndorff1997normal, blattberg2010comparison, laherrere1998stretched, Malevergne2005empirical}, each having their own capabilities, limitations, and complexity level. 

The distributions of returns defined over time scales ranging from 15 s to a few days can often be adequately fitted with power laws~\citep{Gabaix2009power, Gopikrishnan1998inverse}, reflecting their characteristic fat tails for these relatively short time scales. Alternative and more advanced models to fit these distributions include stretched exponentials and log-Weibull distributions~\citep{Malevergne2005empirical}. Return distributions for small values of the time scale have been the subject of extensive data analysis and corresponding modeling efforts. In the light of the forthcoming discussions, we specifically mention Gaussian mixture models that can provide morphological fits of short-term return distributions with asymmetric and/or multimodal shapes~\citep{wirjanto2009applications, cuevas2017gaussian, buckley2008portfolio} and asymmetric L\'{e}vy flights that provide morphological fits of credit ratings and firm accounting ratios~\citep{podobnik2011asymmetric}.

Return distributions defined over large time scales of the order of months and years are a rather underexplored topic in the literature~\citep{dutta2023modeling}. It is commonly assumed that for increasing time scales, return distributions converge toward a Gaussian distribution~\citep{Meng2016quantum, Madan1990variance, Kiyono2006criticality, Kiyono2006power, Tuncay2007power}. In this paper, however, it is shown that for time scales approaching hundreds of trading days, the Gaussian behavior rarely emerges. Instead, one frequently observes non-Gaussian shapes, often exhibiting a high level of asymmetry and even bimodality. This behavior over larger time scales is not well captured by existing models.

Studying return distributions over large time scales provides insights into the long-term dynamics of asset price movements. Such an analysis is particularly relevant for option pricing, where the long-term behavior of asset prices plays an important role in valuing derivatives. By identifying patterns in long-term return distributions, this study complements traditional short-term models for return distributions and can deepen our understanding of the probabilistic structure underlying long-term financial decisions.

In a recent study, Zhan et al.~showed that probability distributions of monthly returns for the Chinese CSI 300 index are bimodal for the majority of months from 2009 to 2023~\citep{zhan2025modeling}. They proposed a parsimonious diffusion model for stock price dynamics, which can generate both a mean-reverting price pattern resulting in a unimodal distribution and a momentum-driven pattern resulting in a bimodal distribution. This recent work highlights the interest in financial models that have the flexibility to capture both unimodal and bimodal long-term return distributions.

In this paper, we apply a recently proposed model based on the discrete-time quantum walk algorithm for financial time series and return distributions~\citep{Debacker2025}. This model is used to characterize the observed asymmetry and bimodality in the return distributions of a variety of financial assets. In contrast to a classical random walk, the discrete-time quantum walk features interference effects in the diffusion process, which allows for the generation of bimodal and asymmetric probability distributions. We aim to model the asymmetry and bimodality in the financial return distributions by leveraging the interference effects inherent in the quantum walk algorithm. In this respect, it represents one of the simplest possible extensions of a classical diffusion process incorporating quantum interference effects. Introducing decoherence into the quantum walk causes it to collapse to a classical random walk~\citep{Debacker2025, mackay2002quantum, brun2003quantumDecoherentCoins, brun2003quantumManyCoins, romanelli2005decoherence, romanelli2011coinflipping, ishak2021entropy}; thus, the quantum walk model for financial time series can be regarded as an extension of a classical model based on geometric Brownian motion~\citep{ziemann2021physics}.

This paper applies the quantum walk model to financial assets with long-term return distributions that exhibit a clear level of asymmetry or bimodality. These features are poorly described by classical models. In this way, this work should be seen as a further data-driven motivation for the quantum walk based model for return distributions proposed in Ref.~\citep{Debacker2025}. It is important to clarify that modeling asset price evolution using a quantum model does not imply that the dynamics of asset prices should be interpreted as a quantum process. Rather, the model serves as a versatile framework to capture non-Gaussian features of asset price dynamics. In this way, it adds to the diversity of existing methodologies and offers an alternative perspective on modeling financial time series and return distributions. Our work contributes to a growing body of literature applying quantum-inspired methodologies to financial and economic problems~\citep{segal1998black, haven2002discussion, schaden2002quantum, shi2006does, ataullah2009wave, zhang2010quantum, meng2015quantum, khrennikova2016application, Meng2016quantum, gao2017quantum, nasiri2018impact, arraut2019connection, orrell2020quantumsupply, sarkissian2020quantum, bhatnagar2022quantum, gomez2022survey, ahn2024business, pires2024synthetic}. This approach is conceptually similar to quantum cognition theory, which likewise does not assume that the brain functions as a quantum computer, but employs concepts from quantum theory such as interference to enrich the formalism and account for phenomena that classical probabilistic frameworks cannot capture~\citep{busemeyer2015quantum, pothos2022quantum}.

In the context of decision theory, discrete-time quantum walks have been employed as a parsimonious probabilistic framework capable of reproducing non-classical behavioral effects through interference~\citep{chen2022use}. The approach combines a compact model formulation and a high degree of generative richness beyond classical models. Similarly, this work exploits the expressivity of the quantum walk algorithm and its associated interference effects to model asymmetric and bimodal financial return distributions. In our quantum walk model, interference effects give rise to bimodality and asymmetry in the return distributions, analogous to their role in quantum-inspired models of decision theory where they produce systematic deviations from classical probability theory. Thus, the interference effects in the quantum walk provide an effective mechanism through which bimodality and asymmetry, as observed in financial market dynamics, emerge at the distributional level. The internal discrete spin degrees of freedom of the quantum walk can be interpreted as encoding directional tendencies in price changes, in analogy with their role in quantum-inspired models of decision theory. A detailed discussion of the interpretation of the quantum walk parameters in a financial context is given in Ref.~\citep{Debacker2025} (see Table~1 therein for an overview).

In the analysis presented here, we focus on the quantum walk algorithm without decoherence effects, as methodologies to introduce decoherence differ in the way the quantum walk distributions gradually evolve toward those of a classical random walk for increasing decoherence rates~\citep{mackay2002quantum, brun2003quantumDecoherentCoins, brun2003quantumManyCoins, romanelli2005decoherence, romanelli2011coinflipping, ishak2021entropy}. Even for relatively low decoherence rates, the resulting distributions resemble classical ones and the distinctive features of quantum walk distributions are lost. Additionally, quantum walks with decoherence involve stochasticity, for example through random phase noise and probabilistic link disruptions; thus, their resulting probability distributions have a certain level of uncertainty. In contrast, quantum walks without decoherence have a fully unitary time evolution that does not involve stochastic elements and yields fully deterministic probability distributions.

The structure of this paper is as follows. In Sect.~\ref{Section:QuantumWalkModel}, we briefly revise the quantum walk algorithm and the associated model for financial time series and the corresponding return distributions. In Sect.~\ref{Sec:ReturnDistributions}, an analysis of empirical time series of asset prices illustrates the emergence of asymmetry and bimodality in distributions of returns defined over large time scales. Section~\ref{Sec:Fits} discusses how the quantum walk model can account for those distinctive features. In Sect.~\ref{Sec:Methodology}, we outline the methodology to fit empirical return distributions with probability distributions resulting from quantum walks. In Sect.~\ref{Sec:Results}, we present and discuss the results of the fitting procedure. Section~\ref{Sec:Conclusions} summarizes our main conclusions. 

\section{Quantum walk model}
\label{Section:QuantumWalkModel}
\subsection{Discrete-time quantum walk}
Before turning to the quantum walk model for financial time series, we briefly revise the quantum walk algorithm~\citep{aharonov1993quantum, kempe2003quantum, venegas2012quantum}, focusing on the parameters that specify its dynamics and its initial conditions. 

The state of the quantum walk system lies in a Hilbert space $\mathcal{H} \equiv \mathcal{H}_C \otimes \mathcal{H}_P$, where the two-dimensional coin space $\mathcal{H}_C$ is spanned by the two spin-1/2 states 
\begin{equation}
    \left| \uparrow \right\rangle = \begin{pmatrix} 1 \\ 
    0
\end{pmatrix},~ \left| \downarrow \right\rangle = \begin{pmatrix} 0 \\ 
1
\end{pmatrix}~,
\end{equation}
and the position space $\mathcal{H}_P$ is spanned by a discrete set of the position basis states $\{ \vert j \in \mathbb{Z} \rangle \}$. A quantum walk of $n$ discrete time steps corresponds to applying the operator $\widehat{V}^n$ to a given initial state $\vert \psi (n=0) \rangle$. The unitary operator $\widehat{V}$ is defined as 
\begin{equation}    \label{EqOperotarV}
   \widehat{V} = \widehat{T} \cdot (\widehat{C} \otimes \widehat{\mathbb{I}}_{P})~,
\end{equation}
where $\widehat{C}$ is the quantum coin toss operator in coin space, $\widehat{\mathbb{I}}_{P}$ is the unity operator in position space, and $\widehat{T}$ is the conditional translation operator. The most general quantum coin toss operator $\widehat{C}$ for a quantum walk initialized at position $j = 0$ is represented by the $2\times2$ matrix~\citep{Debacker2025}
\begin{equation} \label{EqSU2coinDefinition}
    U_{\eta,\theta} = \begin{pmatrix}
  \textrm{e}^{i\eta} \cos{\theta} & \sin{\theta} \\ 
  \sin{\theta} & -\textrm{e}^{-i\eta} \cos{\theta}
\end{pmatrix}~,
\end{equation}
where the angles $\eta \in [0, 2 \pi[$ and $\theta \in [0, \pi/2]$ are referred to as the coin parameters. The conditional translation operator $\widehat{T}$ is defined by
\begin{equation} \label{EqDefinitionTranslation}
    \widehat{T} = \left[ \left| \uparrow \right\rangle \left\langle \uparrow \right| \otimes \left( \sum_{j \in \mathbb{Z}} \vert j + 1 \rangle \langle j \vert \right) \right] + \left[ \left| \downarrow \right\rangle \left\langle \downarrow \right| \otimes \left( \sum_{j \in \mathbb{Z}}  \vert j - 1 \rangle \langle j \vert \right) \right]~,
\end{equation}
and imposes a move to the right (left) for the up (down) component in each time step.

After $n$ time steps, the system's state can be written as 
\begin{equation}   \label{EqWaveFunction}
    \vert \psi (n) \rangle = \widehat{V}^n \vert \psi (n=0) \rangle = \sum_{j = - n}^{ + n} \Bigl( a_j(n) \left| \uparrow \right\rangle + b_j(n) \left| \downarrow \right\rangle \Bigr) \otimes \vert j \rangle ~,
\end{equation}
where the amplitudes $a_j(n)$ and $b_j(n)$ capture the contribution of the up and down components at position $j$. The occupation probability of position $j$ after $n$ discrete time steps is 
\begin{equation}    
\label{EqprobabilityQuantumWalk}
    P_j(n) = | \langle j \vert \psi(n) \rangle |^2 = | a_{j}(n) | ^2 + | b_{j}(n) | ^2 ~.
\end{equation}

Without loss of generality, the Bloch sphere representation can be used for the initial state 
\begin{equation}    \label{EqInitialCondition}
    \vert \psi (n=0) \rangle = \left[ \cos{(\omega / 2)} \left| \uparrow \right\rangle + \exp{\left( i \phi \right)} \sin{(\omega / 2)} \left| \downarrow \right\rangle \right] \otimes \vert j = 0 \rangle ~,
\end{equation}
where $\phi \in [0, 2 \pi[$ and $\omega \in [-\pi, \pi]$ are the initial condition parameters. By selectively tuning the quantum coin parameters ($\eta$, $\theta$) of Eq.~(\ref{EqSU2coinDefinition}) and the initial condition parameters ($\phi$, $\omega$) of Eq.~(\ref{EqInitialCondition}), asymmetric and bimodal probability distributions can be generated~\citep{Debacker2025, chandrashekar2008optimizing}.

In a quantum walk, all possible paths in position space evolve simultaneously, with interference occurring between them. This differs from a classical random walk, where the direction of the next step is decided at each time step. The randomness in a quantum walk only arises upon measurement, while the computation of its probability distribution (Eq.~(\ref{EqprobabilityQuantumWalk})) is fully deterministic.

\subsection{Quantum walk model for financial time series}
In the quantum walk model for financial time series, the asset price increments $\textrm{d}S = S(t + \textrm{d}t) - S(t)$ in a time interval $\textrm{d}t$ are given by~\citep{Debacker2025}
\begin{equation} \label{EqQuantumDiffusionForAssets}
    \textrm{d}S = \mu S(t) \textrm{d}t + \sigma S(t) \bigl[ f(t) \textrm{d}Q(t) \bigr]~,
\end{equation}
where $S(t)$ is the asset price at time $t$, $\mu$ is the rate of a general drift, $\sigma$ is a volatility parameter, $Q(t)$ represents a discrete-time quantum walk process, and $f(t)$ is a function that controls the long-term diffusion properties of the asset price evolution~\citep{Debacker2025}. 

A primary goal of this paper is to use quantum walk probability distributions to model logarithmic return distributions $P_{\Delta t}(g)$ of financial assets. The logarithmic return $g$ of an asset $S$ is defined as the difference in logarithmic price over a duration $\Delta t$
\begin{equation}   \label{EqDefinitionLogReturn}
    g \equiv g(t; \Delta t, S) = \log_{\textrm{e}}{S(t + \Delta t)} - \log_{\textrm{e}}{S(t)}~.
\end{equation}
In the quantum walk model of Eq.~(\ref{EqQuantumDiffusionForAssets}), the drift parameter $\mu$ determines the horizontal shape-preserving translation of the logarithmic return distribution $P_{\Delta t}(g)$. The choice $\mu = 0$ corresponds to an asset price evolution purely governed by the quantum walk process $Q(t)$. When fitting an empirical $P_{\Delta t}(g)$ with a quantum walk distribution, the value of the product $\sigma f(\Delta t)$ can be seen as a rescaling factor for the range over which $g$ extends. Specifically, $\sigma f(\Delta t)$ is chosen to match the scale of the quantum walk fit to that of the empirical return distribution $P_{\Delta t}(g)$. The precise fitting procedure is outlined in Sect.~\ref{Sec:Methodology}. At the end of the following section, the effective long-time scaling of the function $f(t)$ is investigated for a sizeable set of assets.

\section{Identifying asymmetry and bimodality in return distributions}
\label{Sec:ReturnDistributions}
In this section, we discuss the changes in the overall shape of return distributions $P_{\Delta t}(g)$ upon increasing the time scale $\Delta t$. We analyzed the time series of the opening price of 31 financial assets, including stocks and funds. The data were retrieved from \url{https://finance.yahoo.com/} on October 14, 2024. The assets were analyzed over a period of 30 calendar years, or over shorter periods if the data covered less than 30 calendar years. Non-trading days, such as weekend days and public holidays, are excluded from the analysis. Consecutive trading days (for example, a Friday followed by the next Monday) are treated as successive time points in the time series when computing the return distributions. The logarithmic returns are computed using overlapping windows, scanned sequentially along the price time series in order to obtain a smooth probability distribution. Consequently, the resulting return samples are correlated. Table~\ref{table:data} in the appendix provides an overview of all analyzed assets, presenting their ticker, full name, sector, industry, and the end and start date of the time series.

The assets were selected based on the criterion that they exhibit pronounced asymmetry or bimodality in their return distribution at large $\Delta t$. We refer to time scales $\Delta t$ as ``large'' from 100 trading days onwards. To investigate return distributions with $\Delta t \gtrsim 100$ trading days, and to mitigate the impact of correlations arising from overlapping windows, only assets with sufficiently long and continuous price histories can be included in the analysis, thereby excluding assets that have only been on the market for a limited time. Additionally, we focus on assets that exhibit a pronounced level of bimodality or asymmetry in their return distributions at large time scales $\Delta t$. These selection criteria narrow the set of assets included in the analysis.

\begin{figure}[htbp]
\begin{center}
\includegraphics[width=0.5\textwidth,height=2cm,keepaspectratio,angle=0,scale=8.5]{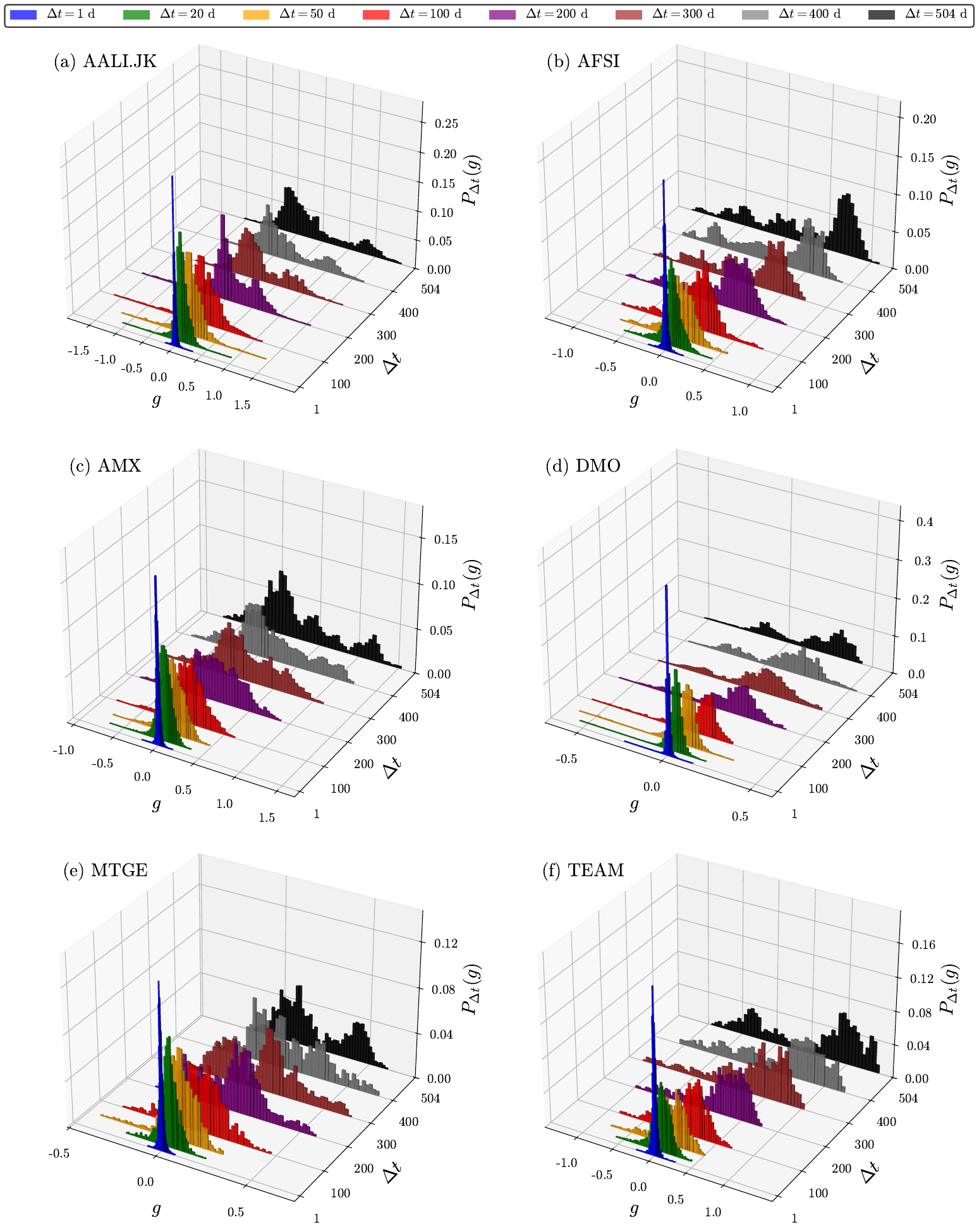}
\caption{\em Waterfall plots of the return distribution $P_{\Delta t}(g)$ of six selected assets for increasing values of the time scale $\Delta t$ (in trading days). The selection of the assets shown here (details in Table~\ref{table:data}) is based on $P_{\Delta t = 504}(g)$ exhibiting a significant degree of bimodality}
\label{FigureWaterfallPlotBimodalDistributions}
\end{center}
\end{figure}

Figure~\ref{FigureWaterfallPlotBimodalDistributions} displays waterfall plots of the $P_{\Delta t}(g)$ of six assets for increasing values of the time scale $\Delta t$. We restrict ourselves to $\Delta t \leq 504$ trading days as this corresponds to two trading years (one trading year equals 252 trading days). For $\Delta t = 1$ trading day, the distributions are sharply peaked and leptokurtic, which is a well-documented finding in the literature~\citep{kou2002jump, hanson2002jump, cont1997scaling, cont2001empirical, liu2022precision, Campbell1998econometrics, Mantegna2000Introduction}. For $\Delta t = 50$ trading days, the distributions transition toward a Gaussian-like shape. For $\Delta t > 50$ trading days, the $P_{\Delta t}(g)$ in Fig.~\ref{FigureWaterfallPlotBimodalDistributions} display clear bimodality, in particular for $\Delta t = 504$ trading days.

\begin{figure}[htbp]
\begin{center}
\includegraphics[width=0.5\textwidth,height=2cm,keepaspectratio,angle=0,scale=4]{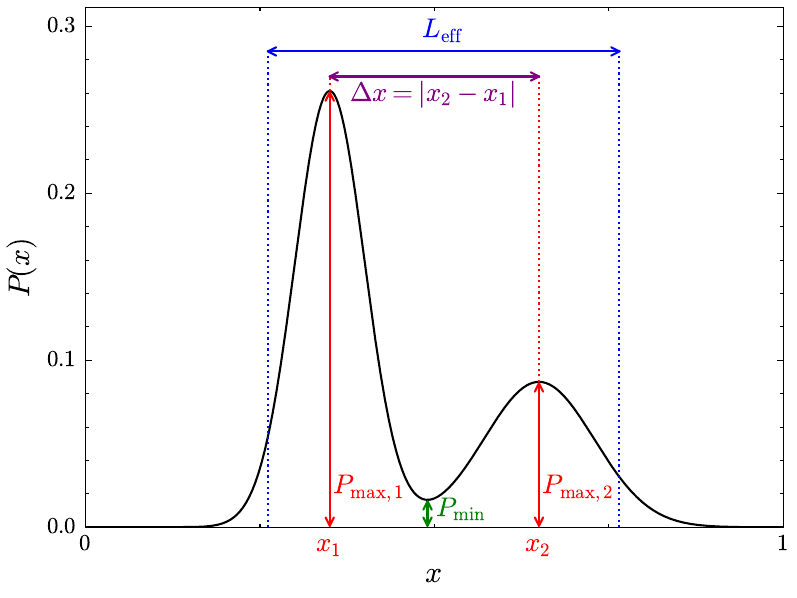}
\caption{\em Illustration of the variables $P_{\textnormal{max,1}}$, $P_{\textnormal{max,2}}$, $P_{\textnormal{min}}$, $\Delta x$, and $L_{\textnormal{eff}}$ contained in the definition of the bimodality measure BM in Eq.~(\ref{EqBimodalityMeasure})}
\label{FigureIllustrationBimodality}
\end{center}
\end{figure}

Several statistical tests have been proposed to discriminate between a unimodal and multimodal distribution~\citep{silverman1981using, hartigan1985dip, muller1991excess, siffer2018your, gupta2024dat}. Rather than merely determining whether a distribution $P(x)$ is bimodal or not, we wish to quantify the degree of bimodality of $P(x)$. Therefore, we introduce a bimodality measure $0 \leq \text{BM} \leq 1$ that satisfies the following two criteria. First, it should adopt a large value if the two peaks are equally high, and if the probability distribution substantially drops between them. Second, $\text{BM}$ should increase for larger separations between the two peaks. A definition that satisfies these criteria is (see Fig.~\ref{FigureIllustrationBimodality} for an illustration of the notation conventions)
\begin{equation}    \label{EqBimodalityMeasure}
    \text{BM} \equiv \left( \frac{P_{\text{max,2}} - P_{\text{min}}}{P_{\text{max,1}}} \right) \times \left( \frac{\Delta x}{L_{\text{eff}}} \right),
\end{equation}
where $P_{\text{max},1}$ and $P_{\text{max},2}$ are the probabilities at the modes (with $P_{\text{max},1} \geq P_{\text{max},2}$), $P_{\text{min}}$ is the minimal value of $P(x)$ between these two modes, $\Delta x$ is the distance between the two modes, and the effective range $L_{\text{eff}}$ is defined as the range that contains the central 95\% of the probability, excluding the outermost 2.5\% on each side of the distribution. The introduction of $L_{\text{eff}}$ in Eq.~(\ref{EqBimodalityMeasure}) guarantees that the value of BM is not heavily impacted by the extreme events on either side of the distribution. As required, the first factor in Eq.~(\ref{EqBimodalityMeasure}) favors an equal height of the two peaks and a substantial drop in probability between them. The second factor favors a larger separation of the two peaks, corresponding to a larger contrast between the events associated with the two modes (two well-separated classes of events). Note that the measure satisfies $0 \leq \text{BM} \leq 1$, except in some pathological distributions where $\Delta x > L_{\text{eff}}$.

\begin{figure}[htbp]
\begin{center}
\includegraphics[width=0.5\textwidth,height=2cm,keepaspectratio,angle=0,scale=4]{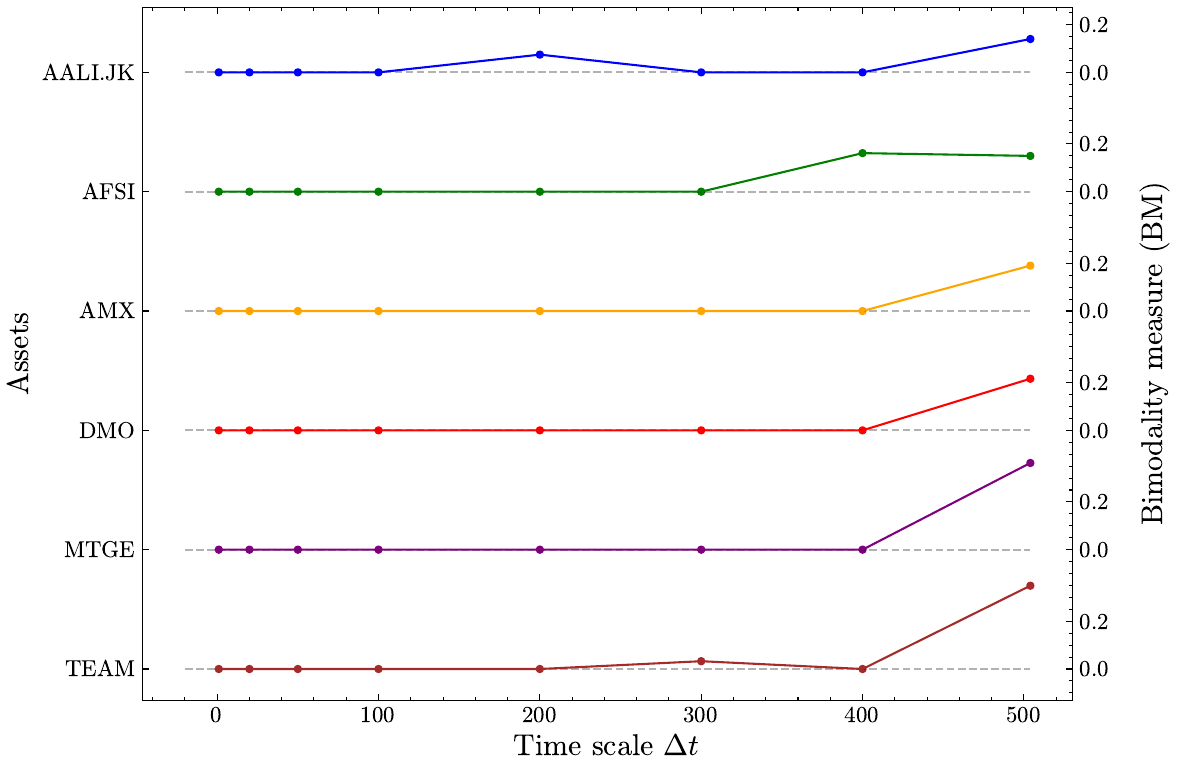}
\caption{\em The $\Delta t$-dependence of the bimodality measure BM (Eq.~(\ref{EqBimodalityMeasure})) of the return distributions $P_{\Delta t}(g)$ of Fig.~\ref{FigureWaterfallPlotBimodalDistributions}}
\label{FigureBimodalityEvolution}
\end{center}
\end{figure}

Figure~\ref{FigureBimodalityEvolution} displays the bimodality measure BM of Eq.~(\ref{EqBimodalityMeasure}) as a function of the time scale $\Delta t$ of the return distributions for the six assets in Fig.~\ref{FigureWaterfallPlotBimodalDistributions}. Upon determining the bimodality measure, we ensure that the global structure of the distribution is captured by binning $P_{\Delta t}(g)$ into about 20 bins to cover the full range of the returns, thereby maintaining a low resolution that prevents small local peaks from emerging. For distributions with only one local maximum, the bimodality is set to zero. For $\Delta t = 2$ trading years, the distributions tend to have a high value of the bimodality measure defined in Eq.~(\ref{EqBimodalityMeasure}). For example, the $P_{\Delta t = 504}(g)$ of TEAM (Fig.~\ref{FigureWaterfallPlotBimodalDistributions}(f)) has two classes in its long-term returns. Indeed, the bump at $g \approx 1$ corresponds with a substantial gain over two trading years, whereas the bump at $g \approx -0.5$ can be associated with a substantial loss.

\begin{figure}[htbp]
\begin{center}
\includegraphics[width=0.5\textwidth,height=2cm,keepaspectratio,angle=0,scale=8.5]{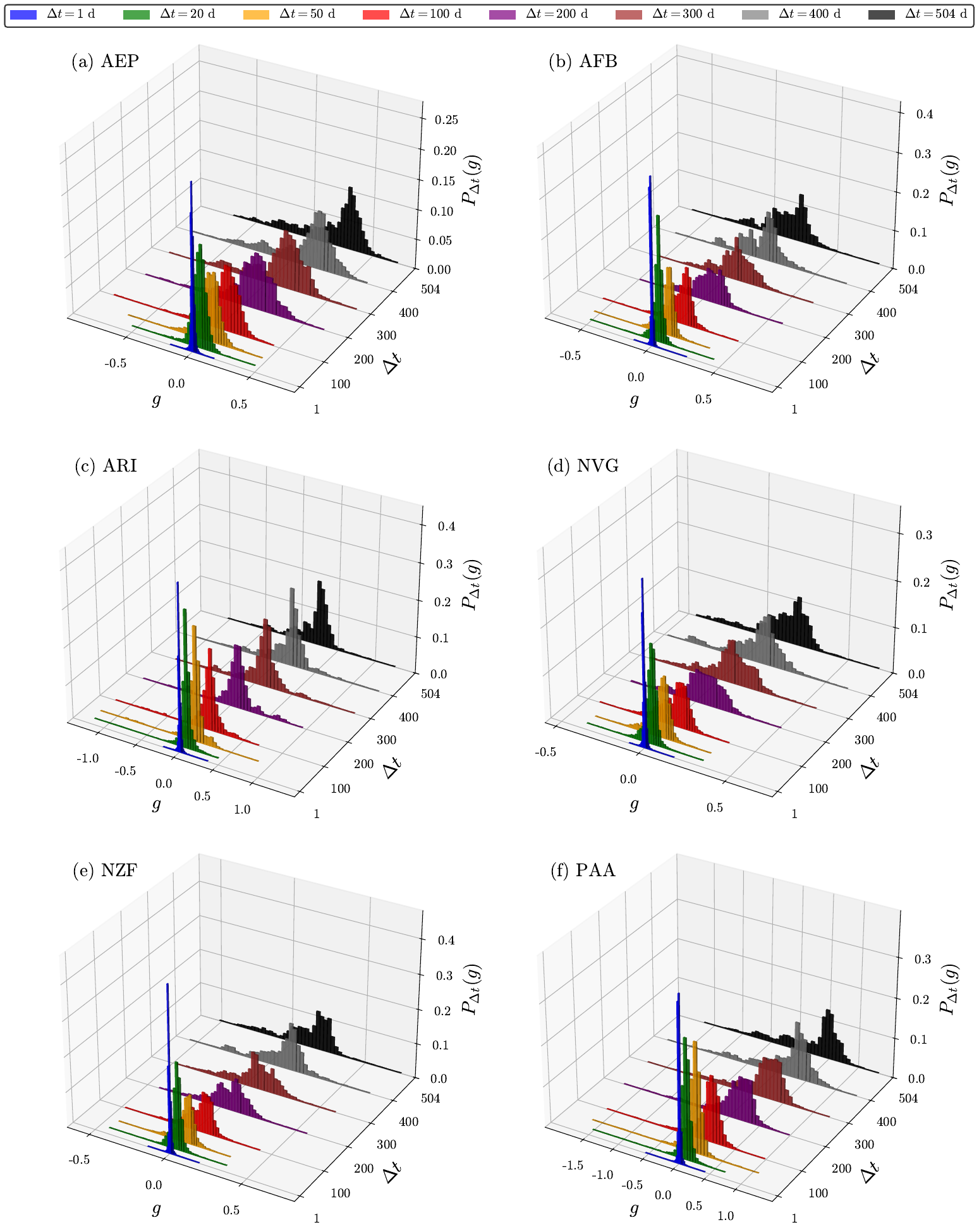}
\caption{\em Waterfall plots of the return distribution $P_{\Delta t}(g)$ of six selected assets for increasing values of the time scale $\Delta t$ (in trading days). The selection of the assets shown here (details in Table~\ref{table:data}) is based on $P_{\Delta t = 504}(g)$ exhibiting a significant degree of asymmetry}
\label{FigureWaterfallPlotSkewedDistributions}
\end{center}
\end{figure}

In Fig.~\ref{FigureWaterfallPlotSkewedDistributions}, we present a similar waterfall plot as in Fig.~\ref{FigureWaterfallPlotBimodalDistributions} for a different set of six selected assets. The return distributions with $\Delta t = $ 504 trading days are non-Gaussian and have a pronounced level of asymmetry, albeit without being bimodal. For $\Delta t \leq 50$ trading days, the trends are similar to those observed in Fig.~\ref{FigureWaterfallPlotBimodalDistributions}. 

\begin{figure}[htbp]
\begin{center}
\includegraphics[width=0.5\textwidth,height=2cm,keepaspectratio,angle=0,scale=4]{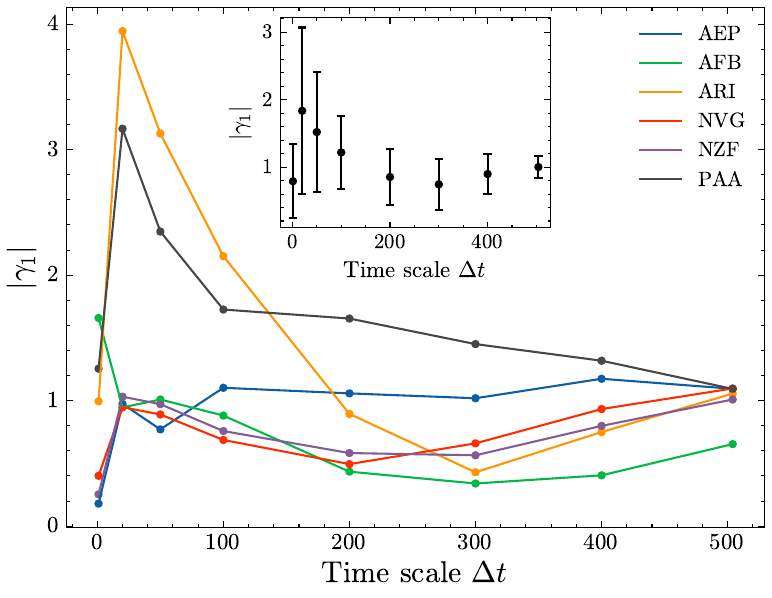}
\caption{\em The $\Delta t$-dependence of the absolute value of the skewness of the return distributions $P_{\Delta t}(g)$ of Fig.~\ref{FigureWaterfallPlotSkewedDistributions}. The inset figure shows the mean and standard deviation of $|\gamma_1|$ across the six assets}
\label{FigureSkewnessEvolution}
\end{center}
\end{figure}

The level of asymmetry in a probability distribution can be quantified by the skewness
\begin{equation}
    \gamma_1 = \frac{\kappa_3}{\kappa_2^{3/2}}~,
\end{equation}
where $\kappa_i$ is the $i$th cumulant. In Fig.~\ref{FigureSkewnessEvolution}, the $|\gamma_1|$ of the return distributions in Fig.~\ref{FigureWaterfallPlotSkewedDistributions} is plotted as a function of the time scale $\Delta t$. We find that the distributions have a persistent level of asymmetry $|\gamma_1| \gtrsim 0.5$ across different time scales $\Delta t$.

Figures~\ref{FigureWaterfallPlotBimodalDistributions}, \ref{FigureBimodalityEvolution}, \ref{FigureWaterfallPlotSkewedDistributions} and \ref{FigureSkewnessEvolution} show that return distributions for $\Delta t \gtrsim 100$ trading days can deviate significantly from a Gaussian profile. Nevertheless, many models relying on long-term predictions of asset prices use a Wiener process or classical random walk as the underlying stochastic mechanism, inherently assuming Gaussian-distributed returns~\citep{ziemann2021physics}. This assumption is exemplified in the Black-Scholes model for financial option pricing~\citep{BlackScholes1973pricing}. These models fail to adequately capture pronounced asymmetry and bimodality in long-term return distributions.

\begin{figure}[htbp]
\begin{center}
\includegraphics[width=0.5\textwidth,height=2cm,keepaspectratio,angle=0,scale=2.7]{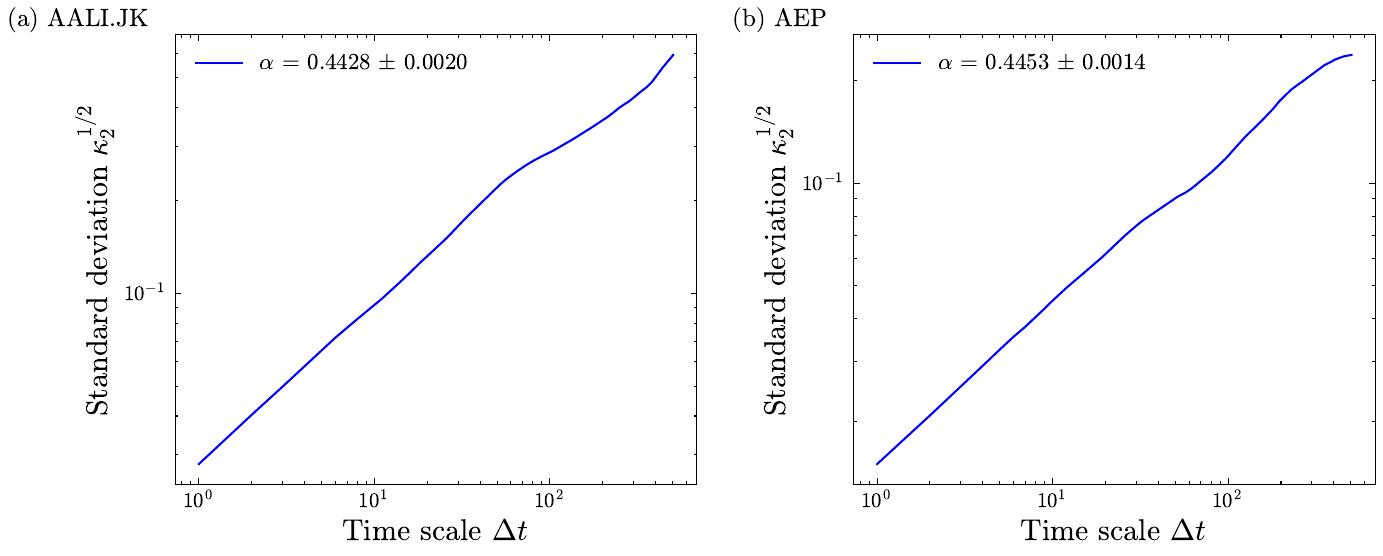}
\caption{\em The $\Delta t$-dependence of the standard deviation $\kappa_2^{1/2}$ of the return distributions $P_{\Delta t}(g)$ for the assets (a) AALI.JK and (b) AEP}
\label{FigureStandardDeviation}
\end{center}
\end{figure}

To conclude this section, we investigate the long-time scaling of the function $f(t)$ in the quantum walk model (Eq.~(\ref{EqQuantumDiffusionForAssets})) that controls the long-term diffusion properties. For empirical guidance in selecting $f(t)$, it is analyzed how the standard deviation $\kappa_2 ^{1/2}$ of the return distributions in Figs.~\ref{FigureWaterfallPlotBimodalDistributions} and \ref{FigureWaterfallPlotSkewedDistributions} scales with $\Delta t$. The standard deviation $\kappa_2 ^{1/2}$ of the logarithmic returns is computed for all time scales $\Delta t$ ranging from 1 to 504 trading days. This is shown for two representative cases in Fig.~\ref{FigureStandardDeviation}. In line with the expectations~\citep{cont2001empirical, dimatteo2007multi}, a power law $\kappa_2 ^{1/2}(\Delta t) \propto (\Delta t) ^{\alpha}$ provides a good fit to the data. The scaling exponent $\alpha$ can be estimated from the linear regression on the logarithmically transformed data. The estimated values of $\alpha$ for the assets from Figs.~\ref{FigureWaterfallPlotBimodalDistributions} and \ref{FigureWaterfallPlotSkewedDistributions} are reported in Table~\ref{table:scalingexponent}. The empirical scaling exponents cluster between 0.43 and 0.54 for most assets in our data set, broadly consistent with an approximate $f(t) \propto t ^{-1/2}$ scaling relation, while a few deviating assets indicate asset-specific differences in the $f(t)$ behavior.

\begin{table}[htbp]
    \centering
    \begin{tabular}{|l|c|}
        \hline
        \textbf{Asset} & \textbf{$\alpha$} \\
        \hline
        AALI.JK & $0.4428 \pm 0.0020$  \\
        AEP & $0.4453 \pm 0.0014$ \\
        AFB & $0.4336 \pm 0.0013$  \\
        AFSI & $0.4963 \pm 0.0013$  \\
        AMX & $0.5350 \pm 0.0015$  \\
        ARI & $0.3422 \pm 0.0045$  \\
        DMO & $0.4599 \pm 0.0028$  \\
        MTGE & $0.4304 \pm 0.0051$  \\
        NVG & $0.4280 \pm 0.0015$  \\
        NZF & $0.4280 \pm 0.0015$  \\
        PAA & $0.4830 \pm 0.0011$  \\
        TEAM & $0.4972 \pm 0.0021$  \\
        \hline
    \end{tabular}
    \caption{The scaling exponent $\alpha$ of the standard deviation $\kappa_2 ^{1/2}$ of the logarithmic returns as a function of the time scale $\Delta t$ for the assets presented in Figs.~\ref{FigureWaterfallPlotBimodalDistributions} and \ref{FigureWaterfallPlotSkewedDistributions}}
    \label{table:scalingexponent}
\end{table}

\section{Quantum walk fits of long-term return distributions}
\label{Sec:Fits}

In Sect.~\ref{Sec:Results}, the return distributions with $\Delta t = 2$ trading years of all 31 assets in the data set analyzed here will be fitted using probability distributions resulting from discrete-time quantum walks. The methodology for this is detailed in the forthcoming subsection.

\subsection{Methodology}     
\label{Sec:Methodology}

For each asset, we first computed the $P_{\Delta t}(g)$ histogram with a small number of bins in order to capture the global structure of the distribution by suppressing the emergence of local spikes. The distributions are classified as unimodal or bimodal based on whether one or two local maxima are detected, adopting the methodology outlined in Sect.~\ref{Sec:ReturnDistributions}. 

A key objective of our analysis is to determine the value of the quantum walk parameters ($\eta$, $\theta$, $\phi$, $\omega$) that yield the best fit to the empirical return distributions. Since the overall shape of a quantum walk probability distribution is not significantly impacted by varying the number of time steps $n$, $n$ is set to 100. Only the width of the $P_j(n)$ of Eq.~(\ref{EqprobabilityQuantumWalk}) is effectively affected by altering $n$, and this can be adjusted for by rescaling the position coordinates. A methodology based on a representative number of time steps $n$ can capture the relevant features of the corresponding quantum walk probability distribution~\citep{Debacker2025}. Moreover, fixing $n$ reduces the number of fitting parameters, which is beneficial for the computational cost of the optimization process. The choice $n=100$ strikes a balance between accuracy and computational cost. Indeed, if $n$ were chosen too small, essential features of the quantum walk distribution could be lost, while a larger $n$ would increase the computational cost of computing the quantum walk (which scales as $\mathcal O (n^2)$) without introducing additional relevant features. In Sect.~\ref{Sec:Results}, we discuss how combining probability distributions from normally distributed values of $n$ can further enhance the quantum walk model's performance in fitting return distributions. The robustness of our methodology with respect to the choices made for $n$ will also be discussed in Sect.~\ref{Sec:Results}. Furthermore, quantum walk probability distributions exhibit spiky behavior due to local interference effects, which differs from what is typically observed in empirical data. To smooth the distinctive peaks, we aggregate each three consecutive positions $j$ with non-zero probability in the quantum walk distribution into one single point, yielding a smoother probability distribution that can be used to fit empirical return distributions.

To determine the appropriate number of bins $N_{\text{bins}}$ for the empirical return distribution used in the quantum walk fit, we first select a value, such that, for a bimodal distribution, the two modes of the quantum walk fit align with the modes of the empirical distribution. For a unimodal asymmetric distribution, the mode and the long tail of the quantum walk fit are made to align smoothly with those of the empirical distribution. After selecting $N_{\text{bins}}$, we introduce two hyperparameters to further optimize the fit. The first hyperparameter adjusts the relative shift between the empirical and quantum walk distributions to assess whether this improves the quality of the fit. The second hyperparameter slightly varies the number of bins in the empirical histogram by a few (up to 4 bins) to investigate whether this adjustment enhances the fit performance. Finally, we impose that both the empirical distribution and the quantum walk fit use the same number of bins, for example by applying zero padding if necessary.

The goodness of fit is quantified by the mean absolute error (MAE)
\begin{equation}
    \text{MAE} = \frac{1}{N_\text{bins}} \sum_{i=1}^{N_{\text{bins}}} \left| P_\text{e}(i) - P_{\text{QW}}(i) \right|~,
\end{equation}
where $P_\text{e}(i)$ and $P_{\text{QW}}(i)$ are the probabilities of the empirical return distribution and the quantum walk fit at bin $i$, respectively. Minimizing the MAE amounts to minimizing the surface between the empirical return distribution and the quantum walk fit. We use the MAE rather than the mean squared error (MSE), since the quadratic penalty in the MSE would disproportionately weight bins near the maxima of the distribution, thereby biasing the fit toward peak heights rather than the shape over the total range of return values. We further note that the probability distributions generated by discrete-time quantum walks do not admit a closed-form analytical expression, thereby precluding the use of likelihood-based methods. The search for the parameters ($\eta$, $\theta$, $\phi$, $\omega$) that minimize the MAE is performed using steepest-ascent hill climbing, which is a local search optimization algorithm~\citep{russell2010artificial}. The process begins with a large set of initial parameter guesses that cover the parameter space with low resolution. The parameter set yielding the lowest MAE becomes the current estimate. In each iteration, the algorithm examines parameter values near the current estimate, progressively narrowing the search space for the optimal parameter values. The current estimate is updated at each iteration to reflect the best-performing parameter set. After a sufficient number of iterations, the optimal values of ($\eta$, $\theta$, $\phi$, $\omega$) are identified to a precision of 0.0001 radians.

In the current analysis, we exclude decoherence effects from our quantum walk model. Our main motivation is that there is no unique way of introducing decoherence effects into the quantum walk algorithm~\citep{mackay2002quantum, brun2003quantumDecoherentCoins, brun2003quantumManyCoins, romanelli2005decoherence, romanelli2011coinflipping, ishak2021entropy}. Examples of methodologies to introduce decoherence are the stochastic disruption of links in the one-dimensional grid and the introduction of a random phase in the quantum coin~\citep{Debacker2025}. The different approaches vary in how they cause the quantum walk distributions to gradually resemble those of a classical random walk as decoherence increases. Moreover, even at low levels of decoherence, the distributions already resemble classical ones, leading to the loss of the characteristic quantum walk features. Finally, due to the stochasticity introduced by decoherence, the resulting probability distributions cannot be determined exactly.

\subsection{Results}     
\label{Sec:Results}

\begin{figure}[htbp]
\begin{center}
\includegraphics[width=0.5\textwidth,height=2cm,keepaspectratio,angle=0,scale=7]{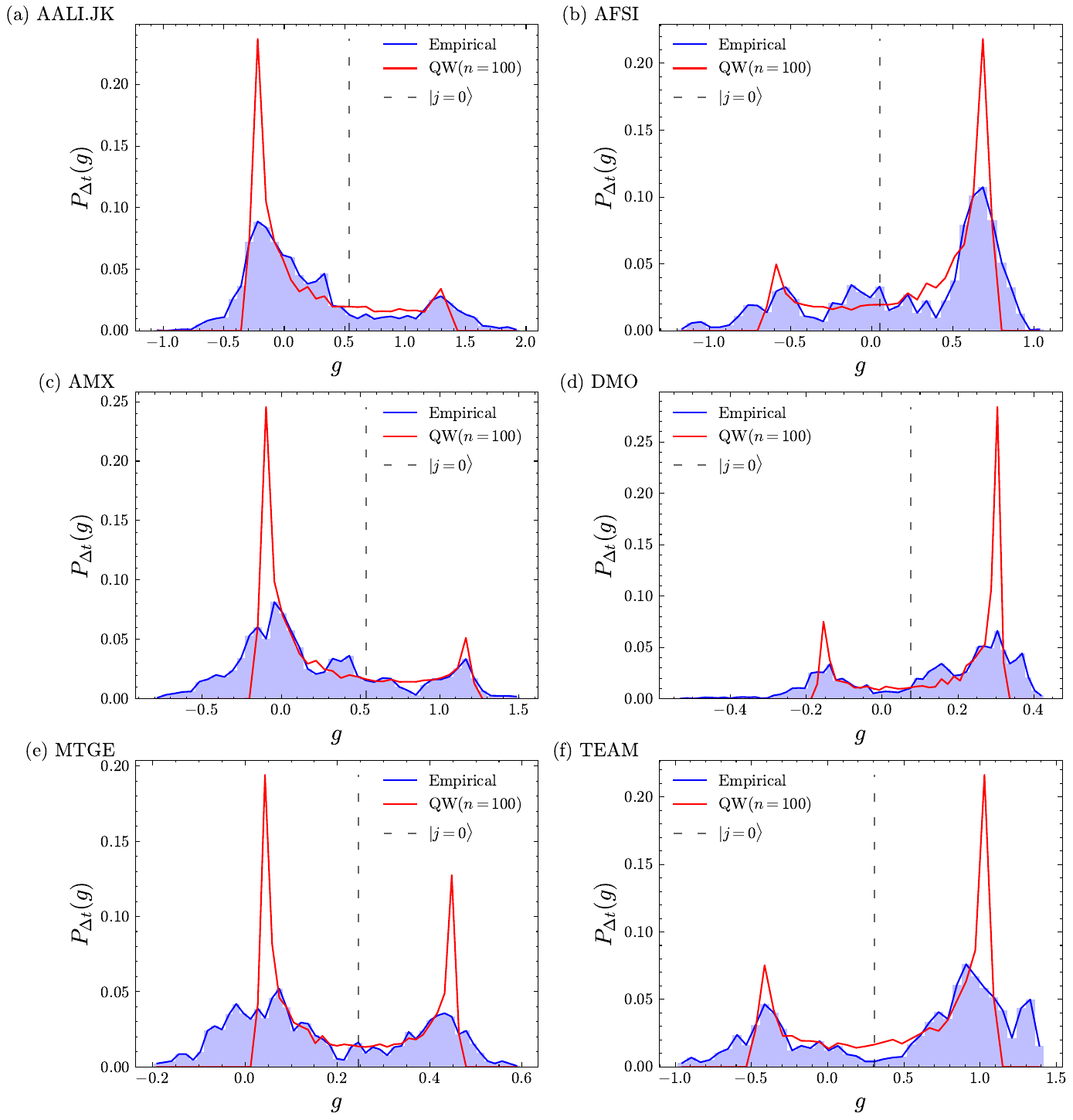}
\caption{\em Quantum walk fit of the bimodal logarithmic return distribution $P_{\Delta t}(g)$ (with $\Delta t = 2$ trading years) for the six selected assets of Fig.~\ref{FigureWaterfallPlotBimodalDistributions}. The number of time steps is set to $n=100$. Both the fit and empirical distribution are normalized. The vertical dashed line corresponds to the initial position $\vert j=0 \rangle$ of the quantum walk}
\label{FigureQuantumWalkFitBimodal}
\end{center}
\end{figure}

\begin{figure}[htbp]
\begin{center}
\includegraphics[width=0.5\textwidth,height=2cm,keepaspectratio,angle=0,scale=7]{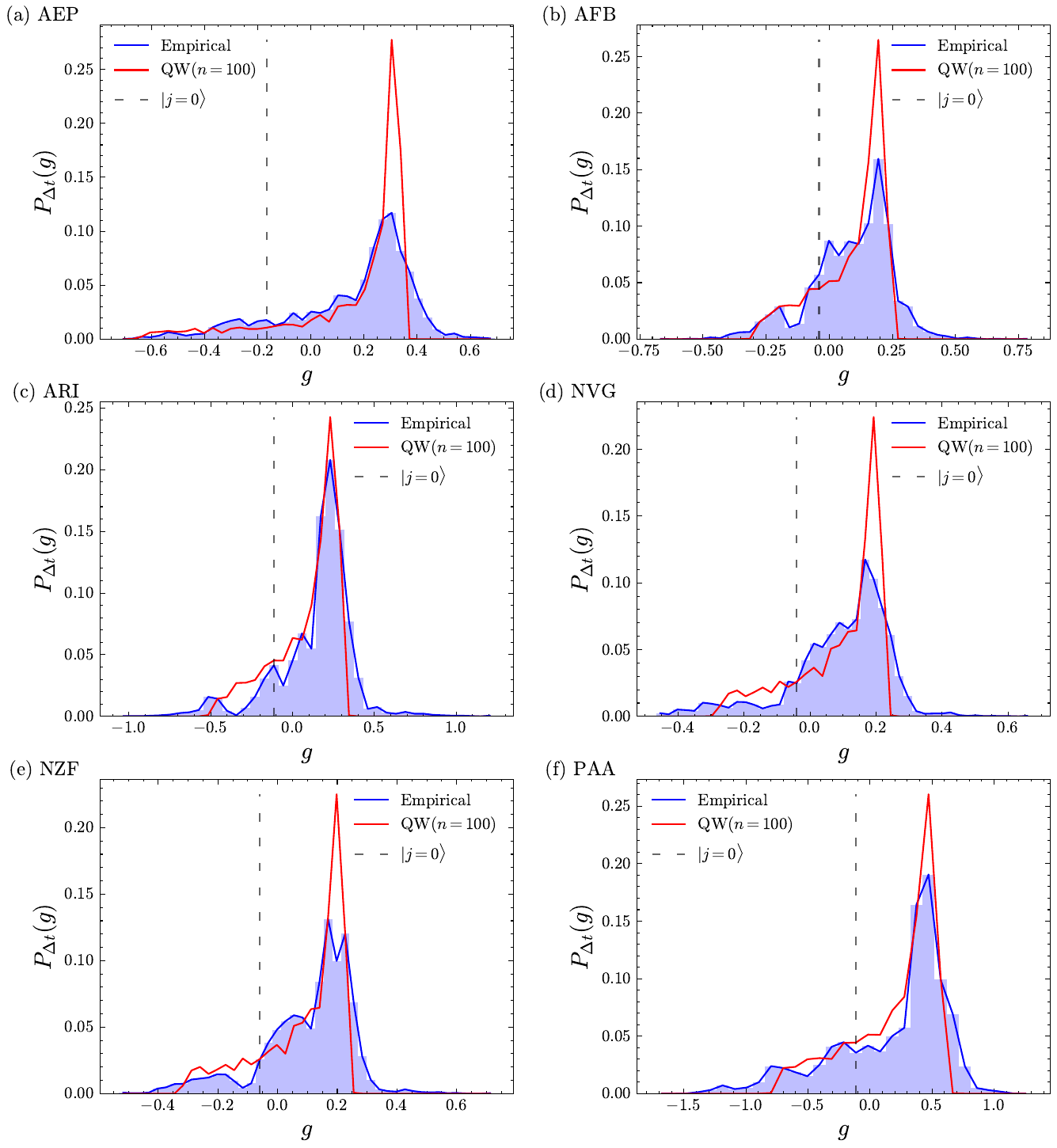}
\caption{\em Quantum walk fit of the logarithmic return distribution $P_{\Delta t}(g)$ exhibiting considerable asymmetry (with $\Delta t = 2$ trading years) for the six selected assets of Fig.~\ref{FigureWaterfallPlotSkewedDistributions}. The number of time steps is set to $n=100$. Both the fit and empirical distribution are normalized. The vertical dashed line corresponds to the initial position $\vert j=0 \rangle$ of the quantum walk}
\label{FigureQuantumWalkFitSkewed}
\end{center}
\end{figure}

In this subsection, we present the quantum walk fits for a broad selection of long-term empirical return distributions $P_{\Delta t}(g)$. In Fig.~\ref{FigureQuantumWalkFitBimodal}, the bimodal return distributions of Fig.~\ref{FigureWaterfallPlotBimodalDistributions} with $\Delta t = 2$ trading years are fitted using the methodology outlined in Sect.~\ref{Sec:Methodology}. In Fig.~\ref{FigureQuantumWalkFitSkewed}, the quantum walk fits of the asymmetric return distributions of Fig.~\ref{FigureWaterfallPlotSkewedDistributions} with $\Delta t = 2$ trading years are displayed. To keep the discussion concise, we highlight these 12 assets, while all remaining ones are succinctly discussed further below.

Although the fits show limitations, they capture the overall behavior of the distributions noticeably better than a Gaussian distribution. For $\Delta t$ of the order of trading years, highly accurate fits are neither a realistic expectation nor the primary goal. Instead, the focus is on systematically capturing the key features of the distributions. The quantum walk fits tend to overestimate the height of the peaks, and sharply fall to nearly zero beyond them. As will be explained below, these issues can be addressed by a further adjustment in the handling of the number of time steps $n$ in the fitting procedure.

The location of the initial position $\vert j=0 \rangle$ of the quantum walk (Eq.~(\ref{EqInitialCondition})) in Figs.~\ref{FigureQuantumWalkFitBimodal} and \ref{FigureQuantumWalkFitSkewed} can be associated with the general drift term in the quantum walk model for financial time series (Eq.~(\ref{EqQuantumDiffusionForAssets})). If the initial position is situated at $g \approx 0$, the drift term can be neglected ($\mu \approx 0$) and the quantum walk can describe the entire diffusion process of the asset price. This is a common feature of the asymmetric distributions in Fig.~\ref{FigureQuantumWalkFitSkewed}. However, this conclusion cannot be generalized to the bimodal distributions presented in Fig.~\ref{FigureQuantumWalkFitBimodal}.

The quantum walk parameters ($\eta$, $\theta$, $\phi$, $\omega$) fix the dynamics and the initial condition of the quantum walk, rather than serving merely as morphological fit parameters of the $P_j(n=100)$ of Eq.~(\ref{EqprobabilityQuantumWalk}). The quantum walk model is suited for modeling bimodal return distributions, as the algorithm can accommodate the dynamics of two competing driving forces modeled by the quantum coin of Eq.~(\ref{EqSU2coinDefinition}) and the conditional translation operator of Eq.~(\ref{EqDefinitionTranslation}). A similar process can be observed in the asset price evolution in Fig.~\ref{FigureWaterfallPlotBimodalDistributions}, where the price typically either rises or falls significantly over a large time interval $\Delta t$. These opposing tendencies contribute to the observed bimodality, making the quantum walk a compelling framework for describing this type of asset price behavior. 

Studying long-term return distributions is highly relevant for risk analysis, as it captures non-trivial characteristics that arise over extended time horizons. In particular, providing a good quantitative estimate of the pronounced negative tails of the empirical distributions in Fig.~\ref{FigureQuantumWalkFitSkewed} and the negative bumps of the distributions in Fig.~\ref{FigureQuantumWalkFitBimodal} is essential, as these relate to projected substantial losses. 

\begin{figure}[htbp]
\begin{center}
\includegraphics[width=0.5\textwidth,height=2cm,keepaspectratio,angle=0,scale=8]{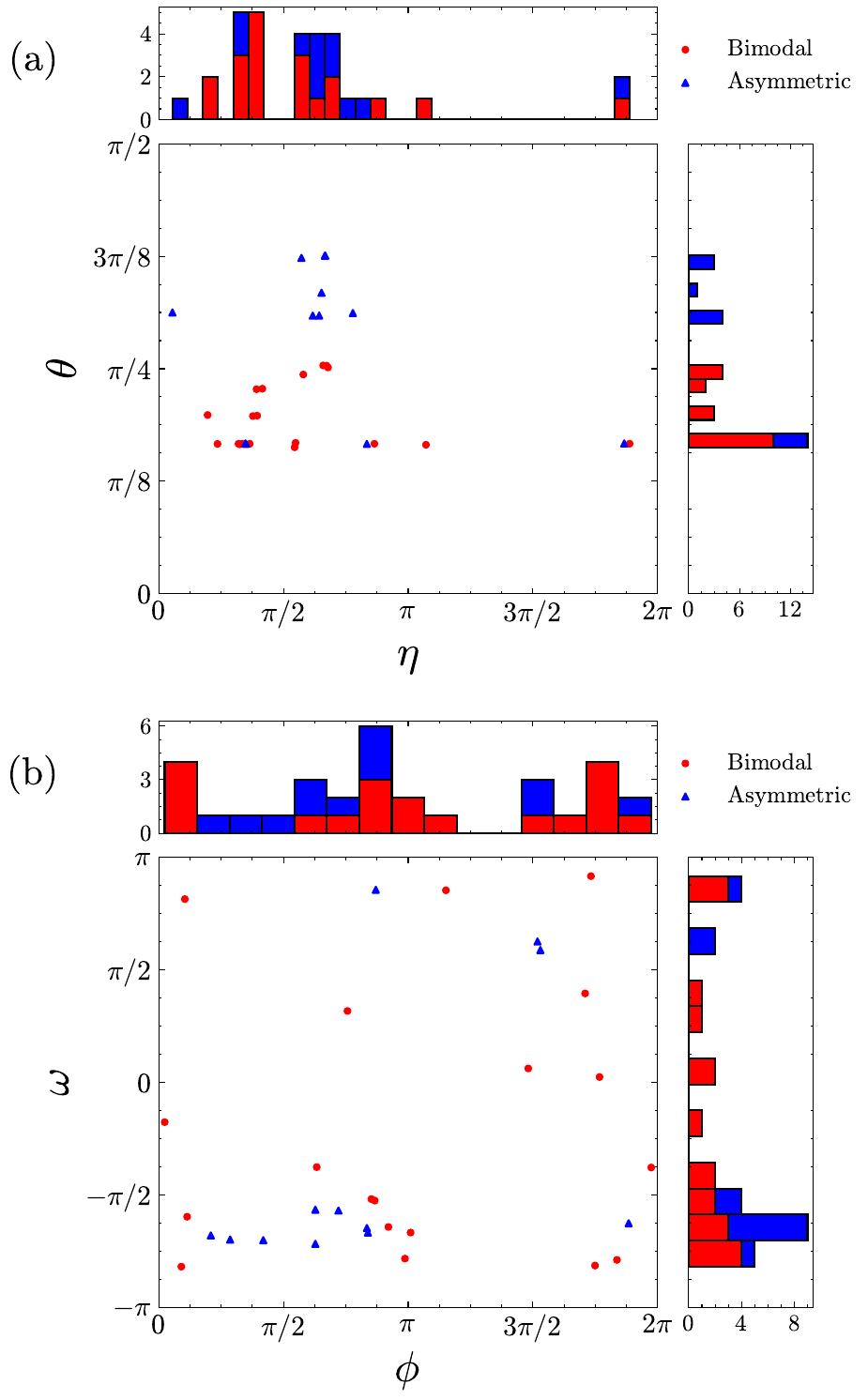}
\caption{\em Scatter plots with marginal histograms of the fitted (a) quantum coin parameters ($\eta$, $\theta$) and (b) initial condition parameters ($\phi$, $\omega$) for all 31 analyzed assets ($\Delta t = 2$ trading years). Bimodal and asymmetric return distributions are distinguished using different marker styles}
\label{FigureParameterSpaceHistograms}
\end{center}
\end{figure}

The scatter plots in Fig.~\ref{FigureParameterSpaceHistograms} display the fitted quantum coin parameters ($\eta$, $\theta$) and initial condition parameters ($\phi$, $\omega$) for all 31 investigated assets (see Table~\ref{table:data} for details), enabling the identification of regions of interest within the parameter space. For reference, limiting situations of the coin operator of Eq.~(\ref{EqSU2coinDefinition}) include the Hadamard coin with ($\eta = 0$, $\theta = \pi / 4$), the $\sigma_z$ Pauli matrix with ($\eta = 0$, $\theta = 0$), and the $\sigma_x$ Pauli matrix with ($\eta = 0$, $\theta = \pi/2$). Upon adopting the $\sigma_z$ and $\sigma_x$ as coin operator, the quantum walk model does not provide realistic return distributions, as reflected by the absence of points near these parameter values in the scatter plot. The coin parameter $\eta$ consistently takes non-zero values, leading to complex numbers in the coin operator. There is a higher level of clustering in the coin parameter space (Fig.~\ref{FigureParameterSpaceHistograms}(a)) than in the initial condition parameter space (Fig.~\ref{FigureParameterSpaceHistograms}(b)). Since the coin parameters enter the dynamics of the quantum walk evolution at each time step, their clustering may suggest similar properties across the investigated assets. Note that the initial condition parameters do not directly influence the dynamics after the start of the quantum walk. Moreover, the bimodal and asymmetric return distributions occupy different regions in the coin parameter space, linking these regions to specific asset price behavior and distinct types of return distributions. A detailed investigation of the distribution and clustering of the fitted quantum walk parameters in parameter space (e.g., using principal component analysis or clustering algorithms) is a promising direction for future research, but falls outside the scope of the current analysis.

As noted earlier, the quantum walk fits in Figs.~\ref{FigureQuantumWalkFitBimodal} and \ref{FigureQuantumWalkFitSkewed} overestimate the height of the peaks of the $P_{\Delta t}(g)$. Additionally, they abruptly drop to nearly zero beyond the peaks, unlike the empirical data. We explore the effect of smoothing the parameter $n$ to resolve these discrepancies and substantially improve the quality of the quantum walk fits. Instead of taking a fixed number of time steps ($n = 100$), which corresponds to a delta distribution $\delta(n = 100)$, we average over 1000 quantum walk probability distributions where $n$ is sampled from a normal distribution $\mathcal{N}(\mu = 100, \sigma^2 = 15^2)$. In this process, the sampled values of $n$ are rounded to the nearest even integer. All other quantum walk parameters keep their values determined during the fit with $n = 100$. 

\begin{figure}[htbp]
\begin{center}
\includegraphics[width=0.5\textwidth,height=2cm,keepaspectratio,angle=0,scale=7]{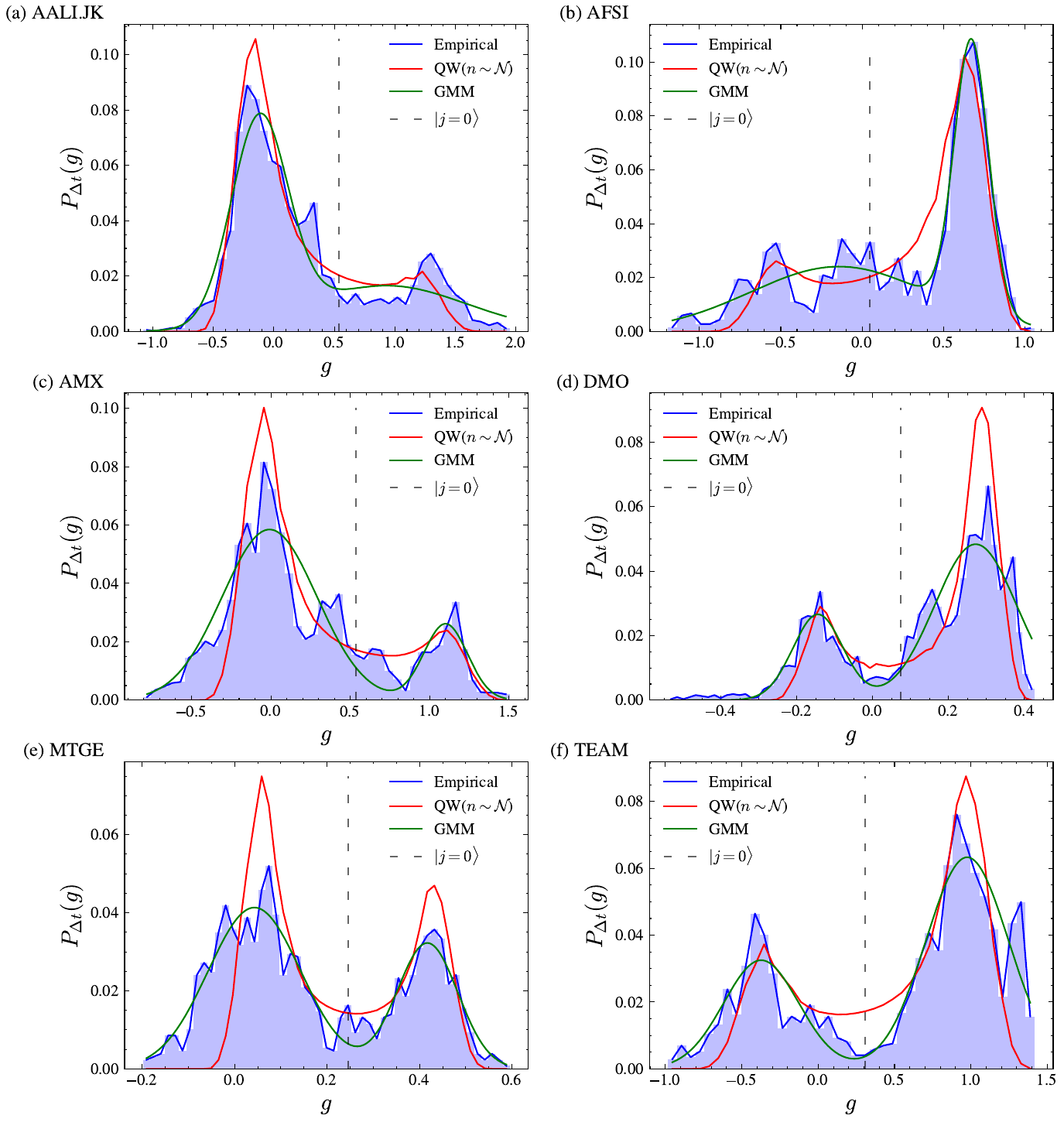}
\caption{\em Quantum walk and two-component GMM fits of the bimodal logarithmic return distribution $P_{\Delta t}(g)$ (with $\Delta t = 2$ trading years) for the six selected assets of Fig.~\ref{FigureWaterfallPlotBimodalDistributions}. The number of time steps $n$ is distributed according to a normal distribution $\mathcal{N}(\mu = 100, \sigma^2 = 15^2)$. The vertical dashed line corresponds to the initial position $\vert j=0 \rangle$ of the quantum walk}
\label{FigureQuantumWalkFitGaussianBimodal}
\end{center}
\end{figure}

\begin{figure}[htbp]
\begin{center}
\includegraphics[width=0.5\textwidth,height=2cm,keepaspectratio,angle=0,scale=7]{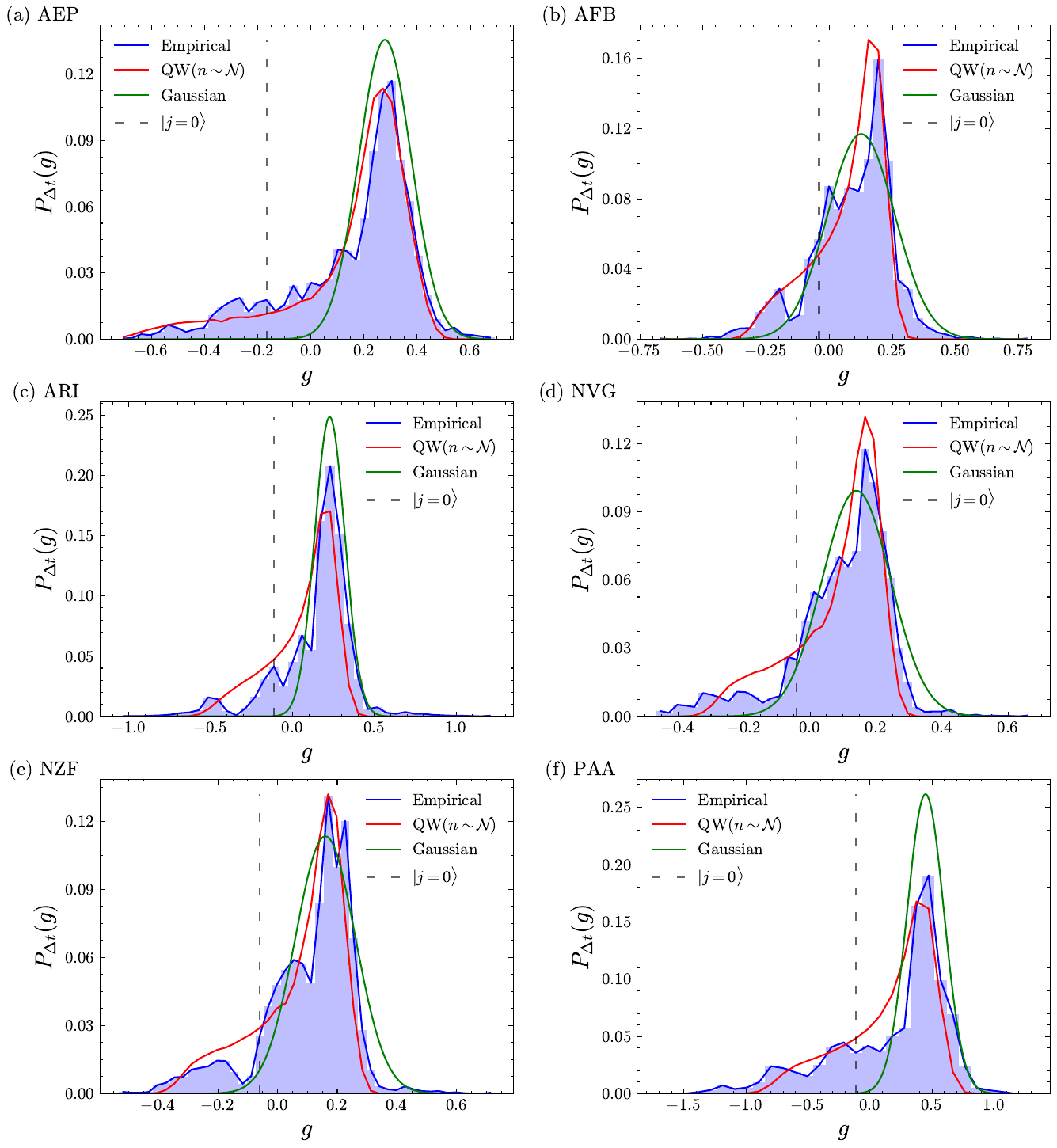}
\caption{\em Quantum walk and Gaussian fits of the logarithmic return distribution $P_{\Delta t}(g)$ exhibiting considerable asymmetry (with $\Delta t = 2$ trading years) for the six selected assets of Fig.~\ref{FigureWaterfallPlotSkewedDistributions}. The number of time steps $n$ in the quantum walk is distributed according to a normal distribution $\mathcal{N}(\mu = 100, \sigma^2 = 15^2)$. The vertical dashed line corresponds to the initial position $\vert j=0 \rangle$ of the quantum walk}
\label{FigureQuantumWalkFitGaussianSkewed}
\end{center}
\end{figure}

\begin{figure}[htbp]
\begin{center}
\includegraphics[width=0.5\textwidth,height=2cm,keepaspectratio,angle=0,scale=5.2]{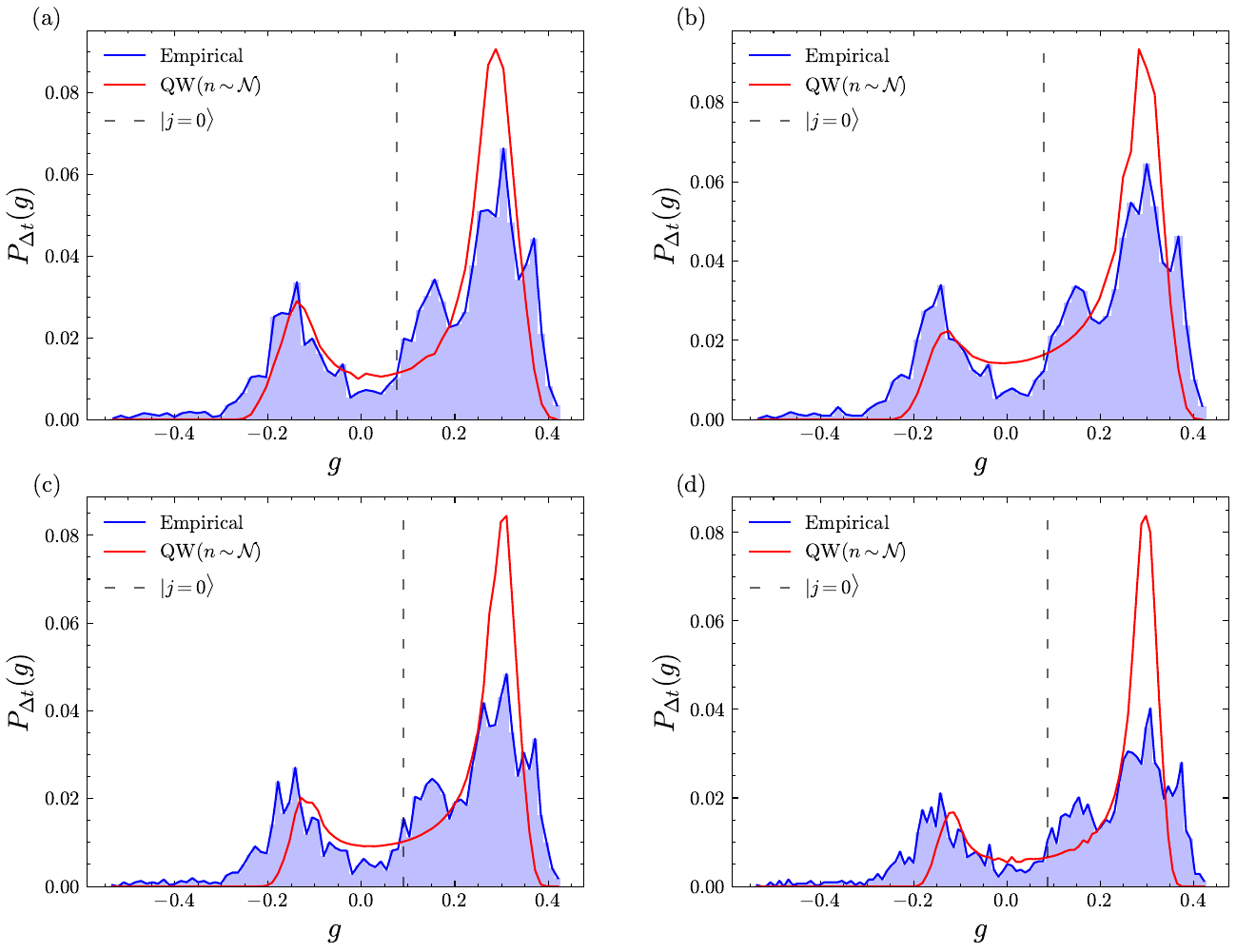}
\caption{\em Quantum walk fit of the logarithmic return distribution $P_{\Delta t}(g)$ (with $\Delta t = 2$ trading years) for the asset DMO. The number of time steps $n$ is distributed according to a normal distribution (a) $\mathcal{N}(\mu = 100, \sigma^2 = 15^2)$, (b) $\mathcal{N}(\mu = 120, \sigma^2 = 18^2)$, (c) $\mathcal{N}(\mu = 140, \sigma^2 = 21^2)$, and (d) $\mathcal{N}(\mu = 160, \sigma^2 = 24^2)$. The vertical dashed line corresponds to the initial position $\vert j=0 \rangle$ of the quantum walk}
\label{FigureInfluenceVaryingN}
\end{center}
\end{figure}

The results of this adapted fitting methodology are displayed in Figs.~\ref{FigureQuantumWalkFitGaussianBimodal} and \ref{FigureQuantumWalkFitGaussianSkewed} for the assets from Figs.~\ref{FigureQuantumWalkFitBimodal} and \ref{FigureQuantumWalkFitSkewed}, respectively. This modification effectively addresses the two issues previously noted. Indeed, the overestimation of peak heights is reduced, and the fitted distributions no longer abruptly drop to zero beyond their peaks, aligning more closely with the behavior of empirical data. Figure~\ref{FigureInfluenceVaryingN} illustrates the effect of varying the mean and standard deviation of the normal distribution used to sample the number of time steps $n$. As a representative case, we consider the asset DMO, which has a bimodal return distribution (panel (d) in Fig.~\ref{FigureQuantumWalkFitGaussianBimodal}). Consistent with earlier discussions, these variations have limited impact on the results of the fitting procedure. As the essential features of the quantum walk fit remain unaffected, this observation supports the use of a fixed distribution $\mathcal{N}(\mu = 100, \sigma^2 = 15^2)$ in the main analysis.

In Fig.~\ref{FigureQuantumWalkFitGaussianSkewed}, we also present a Gaussian fit of the asymmetric return distributions $P_{\Delta t}(g)$. Note that the quantum walk model more effectively captures the general shape of the return distributions compared to the Gaussian distribution. In Fig.~\ref{FigureQuantumWalkFitGaussianBimodal}, we also present a two-component Gaussian mixture model (GMM) fit of the bimodal return distributions $P_{\Delta t}(g)$. In Table~\ref{table:MAE}, we list the MAE and the Kolmogorov–Smirnov (KS) statistic of the quantum walk fits (with $n \sim \mathcal{N}(\mu = 100, \sigma^2 = 15^2)$), the Gaussian fits, and the two-component GMM fits for the distributions shown in Figs.~\ref{FigureQuantumWalkFitGaussianBimodal} and \ref{FigureQuantumWalkFitGaussianSkewed}. For the asymmetric distributions, except for the cases of ARI and NVG, the MAE of the quantum walk fit is lower than for the Gaussian fit, indicating a better characterization of the return distribution. Moreover, across all asymmetric distributions, the KS statistic is consistently lower for the quantum walk fit compared to the Gaussian fit. These results underscore the shortcomings of the classical model and the superior overall performance of the quantum walk model in characterizing asymmetric return distributions. We note, however, that for the bimodal distributions, a two-component GMM can achieve lower MAE and KS values, indicating a better morphological fit to the data. However, the parameters obtained from the GMM (e.g., the relative weight and standard deviations of the Gaussian components) are purely morphological fit parameters. They describe the shape of the distribution but do not carry additional information. In contrast, the parameters of the quantum walk are part of a generative dynamical framework and control the mechanisms through which the resulting return distributions are produced. Moreover, unlike the single Gaussian resulting from geometric Brownian motion, GMMs lack a single straightforward diffusive process as an underlying generative mechanism. Although GMMs can yield decent morphological fits, they fail to provide a dynamical interpretation. A key advantage of the quantum walk model is that it can characterize both unimodal asymmetric and bimodal distributions within a single framework.

\begin{table}[htbp]
    \centering
    \begin{tabular}{|l|ccc|ccc|}
        \hline
         & \multicolumn{3}{c|}{\textbf{MAE} ($10^{-2}$)} & \multicolumn{3}{c|}{\textbf{KS} ($10^{-1}$)} \\
        \textbf{Asset} & \textbf{QW ($\boldsymbol{n \sim \mathcal{N}}$)} & \textbf{Gaussian} & \textbf{GMM} & \textbf{QW ($\boldsymbol{n \sim \mathcal{N}}$)} & \textbf{Gaussian} & \textbf{GMM} \\
        \hline
        AALI.JK & 0.640 & & 0.563 & 0.661 & & 0.305 \\
        AEP & 0.524 & 0.964 & & 0.456 & 2.953 & \\
        AFB & 0.813 & 0.899 & & 0.908 & 1.191 & \\
        AFSI & 0.881 & & 0.526 & 0.835 & & 0.235 \\
        AMX & 0.844 & & 0.649 & 1.241 & & 0.446 \\
        ARI & 1.170 & 0.818 & & 2.089 & 2.368 & \\
        DMO & 0.729 & & 0.435 & 0.997 & & 0.592 \\
        MTGE & 0.906 & & 0.369 & 1.999 & & 0.189 \\
        NVG & 0.757 & 0.745 & & 0.777 & 1.380 & \\
        NZF & 0.774 & 0.918 & & 1.172 & 1.775 & \\
        PAA & 1.209 & 1.452 & & 1.309 & 3.414 & \\
        TEAM & 0.965 & & 0.578 & 1.062 & & 0.206 \\
        \hline
    \end{tabular}
    \caption{MAE and KS statistic of the quantum walk fit (with $n \sim \mathcal{N}(\mu = 100, \sigma^2 = 15^2)$) of the return distributions presented in Figs.~\ref{FigureQuantumWalkFitGaussianBimodal} and \ref{FigureQuantumWalkFitGaussianSkewed}, the Gaussian fit of the asymmetric return distributions (Fig.~\ref{FigureQuantumWalkFitGaussianSkewed}) and the two-component GMM fit of the bimodal return distributions (Fig.~\ref{FigureQuantumWalkFitGaussianBimodal})}
    \label{table:MAE}
\end{table}

The analysis presented in this work is primarily phenomenological, focusing on the empirical shape of long-term return distributions. Nevertheless, it is important to address the interpretation of the observed bimodality and asymmetry. We emphasize that these features are not claimed to arise from genuine quantum interference effects in financial markets. The quantum walk is not intended as a literal microscopic description of market dynamics, but rather as a compact modeling framework capable of encoding interference-like effects emerging in financial data. The most plausible interpretation of the emergence of bimodality at large time scales is the non-stationarity of the financial time series, where returns reflect the mixing of distinct regimes, such as bull and bear markets, crisis periods, or shifts in monetary and economic policy~\citep{hamilton1989new, guidolin2006econometric, ang2012regime}. In this context, the added value of the quantum walk framework lies in its ability to reproduce bimodality and asymmetry within a single coherent model, without requiring explicit identification or labeling of individual regimes. In this sense, the quantum walk can provide a parsimonious description of the emergent statistical features resulting from the superposition of heterogeneous regimes.

The results in Figs.~\ref{FigureQuantumWalkFitGaussianBimodal} and \ref{FigureQuantumWalkFitGaussianSkewed} demonstrate the potential of the quantum walk model to capture key characteristics of long-term return distributions and the diffusion process of asset prices. Discrete-time quantum walks have been proposed earlier in a model for financial option pricing~\citep{orrell2020quantummathsbook, orrell2021quantumwalkArticle, Orrell:MoneyMagic}. In this option pricing model, the probability distributions resulting from a quantum walk with the Hadamard coin and the symmetric initial state $\vert \psi (n=0) \rangle = 1/ \sqrt{2} \left( \left| \uparrow \right\rangle + i \left| \downarrow \right\rangle \right) \otimes \vert j = 0 \rangle$ were used to represent economic agents' projections on the future value of the underlying asset. Our data analysis reinforces the assumptions of this model, as the occasional bimodal characteristics in long-term return distributions are incorporated into the option pricing model. Moreover, the results above illustrate that our quantum walk model offers an adequate characterization of bimodal return distributions, further corroborating the viability of the quantum walk-based option pricing model.

\section{Conclusions}     
\label{Sec:Conclusions}
We analyzed asset price variations over extended periods using a model based on the discrete-time quantum walk, demonstrating its ability to capture specific empirical characteristics more effectively than a Gaussian model. We provided evidence for the emergence of bimodality and asymmetry in return distributions defined over time scales exceeding several months. The latter remain a relatively underexplored topic in econophysics and econometrics, despite their importance for models involving long-term predictions of asset prices and long-term risk assessment. Capturing their characteristics is particularly relevant for option pricing models, where accurate representations of return distributions over long time horizons are key to assessing derivative values and associated risks. In particular, a bimodal distribution can be associated with a combination of substantial losses and gains, having significant implications for long-term risk management. We used a quantum walk model to characterize the bimodality and asymmetry in these long-term return distributions, leveraging the quantum interference effects intrinsic to its dynamics to capture these empirical features. We stress that financial time series are not quantum processes, yet the quantum walk is employed as a modeling tool to capture empirical features, such as asymmetry and bimodality in long-term returns. Although there is still room to improve the quality of the quantum walk fits, they reflect the observed behavior substantially better than a Gaussian distribution. By identifying regions within the coin parameter space that can be associated with bimodal and asymmetric return distributions, we can relate the quantum walk parameters to different classes of long-term asset price behavior.

In future work, the analysis can be extended to larger data sets, and more refined fitting strategies can be explored to further improve the quality of the quantum walk fits. Nevertheless, at large time scales, such as those of two trading years, achieving a perfect characterization of the return distributions is inherently challenging. A larger data set would also enable a more detailed analysis of the distribution and clustering of the fitted quantum walk parameters, for example by examining whether assets from different economic sectors occupy distinct regions of parameter space. Another interesting research question is whether there is a discernible difference in the quantum walk parameters for the return distributions of composite assets and of single assets. The analysis presented in this work contains both classes, but the sample sizes are too small for a detailed comparative analysis. A further line for future research lies in the inclusion of decoherence effects into the quantum walk algorithm to investigate how return distributions evolve from smaller to larger time scales. Indeed, by including decoherence effects, one can generate asymmetric probability distributions with heavier tails than a Gaussian distribution, which offers applications for modeling short-term return distributions. Allowing for decoherence within the quantum walk would further increase the expressivity of the model, enhancing its applicability at the cost of introducing an additional parameter.

\section*{Acknowledgments}
This project was conducted with the support of the Special Research Fund of Ghent University (Projects Nos. BOF/DOC/2023/103 and BOF/BAF/4y/24/1/018).

\section*{Author contributions}
All authors contributed to the study conception and design. Data collection and analysis were performed by Stijn De Backer. The first draft of the manuscript was written by Stijn De Backer and all authors commented on previous versions of the manuscript. All authors read and approved the final manuscript.

\section*{Data availability statement}
The data used in this study are available from Yahoo Finance (\url{https://finance.yahoo.com/}).

\section*{Declaration of conflict of interest}
The authors have no relevant financial or non-financial interests to disclose.

\section*{Accepted manuscript notice}
This version of the article has been accepted for publication, after peer review but is not the Version of Record and does not reflect post-acceptance improvements, or any corrections. The Version of Record is available online at: \url{https://doi.org/10.1140/epjb/s10051-026-01157-8}. Use of this Accepted Version is subject to the publisher's Accepted Manuscript terms of use \url{https://www.springernature.com/gp/open-science/policies/accepted-manuscript-terms}.

\begin{appendices}

\section{Data}\label{secA1}

Table~\ref{table:data} provides an overview of the analyzed assets, presenting their ticker, full name, sector, industry, and the end and start date of the analyzed time series. The data set was obtained from \url{https://finance.yahoo.com/} on October 14, 2024.

\setlength\LTleft{0pt}
\setlength\LTright{0pt}
\begin{longtable}{p{1.5cm} p{3.2cm} p{2.8cm} p{3.2cm} l l}
\hline
\textbf{Ticker} & \textbf{Full name} & \textbf{Sector} & \textbf{Industry} & \textbf{Start date} & \textbf{End date} \\
\hline
\endfirsthead

\hline
\textbf{Ticker} & \textbf{Full name} & \textbf{Sector} & \textbf{Industry} & \textbf{Start date} & \textbf{End date} \\
\hline
\endhead

6599.KL & Aeon Co. (M) Bhd. & Consumer Cyclical & Department Stores & 2000-01-03 & 2024-10-14 \\
AALI.JK & PT Astra Agro Lestari Tbk & Consumer Defensive & Farm Products & 2001-04-05 & 2024-10-14 \\
AAOI & Applied Optoelectronics, Inc. & Technology & Communication Equipment & 2013-09-26 & 2024-10-11 \\
AB & AllianceBernstein Holding L.P. & Financial Services & Asset Management & 1994-10-19 & 2024-10-11 \\
ACV & Virtus Diversified Income and Convertible Fund & Financial Services & Asset Management & 2015-05-21 & 2024-10-11 \\
ADDYY & adidas AG & Consumer Cyclical & Footwear and Accessories & 2006-05-31 & 2024-10-11 \\
ADX & Adams Diversified Equity Fund, Inc. & Financial Services & Asset Management & 1994-10-19 & 2024-10-11 \\
AEP & American Electric Power Company, Inc. & Utilities & Utilities - Regulated Electric & 1994-10-19 & 2024-10-11 \\
AFB & AllianceBernstein National Municipal Income Fund, Inc. & Financial Services & Asset Management & 2002-01-29 & 2024-10-11 \\
AFSI & AmTrust Financial Services, Inc. & Financial Services & Insurance - Property and Casualty & 2006-11-13 & 2018-12-10 \\
AJX & Great Ajax Corp. & Real Estate & REIT - Mortgage & 2015-02-13 & 2024-10-11 \\
AM & Antero Midstream Corporation & Energy & Oil and Gas Midstream & 2017-05-04 & 2024-10-11 \\
AMCX & AMC Networks Inc. & Communication Services & Entertainment & 2011-06-16 & 2024-10-11 \\
AMG & Affiliated Managers Group, Inc. & Financial Services & Asset Management & 1997-11-21 & 2024-10-11 \\
AMX & América Móvil, S.A.B. de C.V. & Communication Services & Telecom Services & 2001-02-12 & 2024-10-11 \\
ANY & Sphere 3D Corp. & Financial Services & Capital Markets & 2013-08-12 & 2024-10-11 \\
ARI & Apollo Commercial Real Estate Finance, Inc. & Real Estate & REIT - Mortgage & 2009-09-24 & 2024-10-11 \\
AWK & American Water Works Company, Inc. & Utilities & Utilities - Regulated Water & 2008-04-23 & 2024-10-11 \\
BABA & Alibaba Group Holding Limited & Consumer Cyclical & Internet Retail & 2014-09-19 & 2024-10-11 \\
DMO & Western Asset Mortgage Opportunity Fund Inc. & Financial Services & Asset Management & 2010-02-24 & 2024-10-11 \\
EVTC & EVERTEC, Inc. & Technology & Software - Infrastructure & 2013-04-12 & 2024-10-11 \\
GBAB & Guggenheim Taxable Municipal Bond and Investment Grade Debt Trust & Financial Services & Asset Management & 2010-10-28 & 2024-10-11 \\
IGI & Western Asset Investment Grade Opportunity Trust Inc. & Financial Services & Asset Management & 2009-07-01 & 2024-10-11 \\
ISL & Aberdeen Israel Fund, Inc. & Financial Services & Asset Management & 2009-09-08 & 2018-04-30 \\
MTGE & MTGE Investment Corp. & Real Estate & REIT - Mortgage & 2011-08-04 & 2018-09-17 \\
NSA & National Storage Affiliates Trust & Real Estate & REIT - Industrial & 2015-04-22 & 2024-10-11 \\
NVG & Nuveen AMT-Free Municipal Credit Income Fund & Financial Services & Asset Management & 2002-09-12 & 2024-10-11 \\
NZF & Nuveen Municipal Credit Income Fund & Financial Services & Asset Management & 2001-10-03 & 2024-10-11 \\
PAA & Plains All American Pipeline, L.P. & Energy & Oil and Gas Midstream & 1998-11-18 & 2024-10-11 \\
SAVE & Spirit Airlines, Inc. & Industrials & Airlines & 2011-05-26 & 2024-10-11 \\
TEAM & Atlassian Corporation & Technology & Software - Application & 2015-12-09 & 2024-10-11 \\
\hline
\caption{Overview of the analyzed assets, including their ticker, full name, sector, industry, and the start and end date of the analyzed time series} 
\label{table:data} \\
\end{longtable}

\end{appendices}


\bibliography{sn-bibliography}

\end{document}